\documentclass[letterpaper, 10pt, journal, onecolumn, final]{IEEEtran}

\IEEEoverridecommandlockouts                              %

\usepackage{cite}
\usepackage{amsmath,amssymb,amsfonts}
\usepackage{graphicx}
\usepackage{algorithm,algorithmicx,algpseudocode}
\usepackage{tcolorbox}
\usepackage{booktabs} %
\usepackage{subcaption}
\usepackage{caption}
\usepackage{soul}

\title{\huge Distance-Certified Safe Control in Unknown Environments \\ via Spatio-Temporal Composition of Gaussian Processes}

\author{Davide Zarabini, Stefano Di Gregorio, Giuseppe Notarstefano
\thanks{The work was partially funded by the FISA-2023-00210 project APACHE - CUP
      J53C25000520001, by Next Generation EU, National Recovery and Resilience Plan (NRRP), Mission
      4, Component 2, Investment 3.3; CUP J33C24001490009, and by IMA
      S.p.A. 
The corresponding author is Davide Zarabini.}
\thanks{The authors are with the Department of
    Electrical, Electronic and Information Engineering, University of
    Bologna, 40136, Bologna, Italy (e-mail:
    davide.zarabini@unibo.it, stefano.digregorio@unibo.it,
    giuseppe.notarstefano@unibo.it).}%
}

\newcommand{\realspace}{\mathbb{R}}
\newcommand{\workspace}{\mathcal{W}}
\newcommand{\obstacle}{\mathcal{O}}
\newcommand{\trueboundary}{\partial\obstacle}

\newcommand{\obspoints}{X_{\nobs}}

\newcommand{\nobs}{N}

\newcommand{\obspoint}[1]{p_{#1}}
\newcommand{\pointsetdist}{\operatorname{dist}}
\newcommand{\hausdorffdist}{d_{\mathrm H}}
\newcommand{\euclidnorm}[1]{\left\lVert#1\right\rVert}
\newcommand{\requiredclearance}{d}

\newcommand{\covererror}{\delta_{\nobs}}
\newcommand{\selectionbeta}{\beta}
\newcommand{\selectionpower}{\nu}
\newcommand{\selectionerror}{\varepsilon_{\selectionbeta,\nobs}}
\newcommand{\sqdist}[1]{q_{#1}}

\newcommand{\weight}[1]{w_{#1}}

\newcommand{\prior}{m_0}
\newcommand{\obslabel}{y_{\mathrm{obs}}}
\newcommand{\signalvariance}{\sigma_f^2}
\newcommand{\noisevariance}{\sigma_n^2}
\newcommand{\postamp}{A}

\newcommand{\gplength}{\ell}

\newcommand{\gpkernel}{k_{\obspoints}}
\newcommand{\gpgram}{K_{\obspoints}}
\newcommand{\gpmean}{\mu_{\obspoints}}

\newcommand{\safetyfun}{H}

\newcommand{\controlstate}{x}
\newcommand{\controlinput}{u}

\newcommand{\controlmap}{P}
\newcommand{\workspacepoint}{p}

\newcommand{\nominalinput}{u_{\mathrm{nom}}}
\newcommand{\safeinput}{u_{\mathrm{safe}}}

\newcommand{\cC}{\mathcal{C}}
\newcommand{\cX}{\mathcal{X}}
\newcommand{\cU}{\mathcal{U}}
\newcommand{\cK}{\mathcal{K}}

\newcommand{\xstar}{x_*}

\newcommand\oprocendsymbol{\hbox{$\square$}}
\newcommand\oprocend{\relax\ifmmode\else\unskip\hfill\fi\oprocendsymbol}
\def\eqoprocend{\tag*{$\square$}}

\newcommand{\cbf}{h}
\newcommand{\R}{\mathbb{R}}

\newcommand{\safesetphys}{\mathcal{S}_d}
\newcommand{\velobsmax}{\bar{v}_o}
\newcommand{\velboundmax}{\bar{v}_\partial}

\newcommand{\controlstrategy}{Distance-Certified GP-based Safe Controller }

\newtheorem{assumption}{Assumption}
\newtheorem{lemma}{Lemma}
\newtheorem{proposition}{Proposition}
\newtheorem{theorem}{Theorem}

\newtheorem{remark}{Remark}
\newtheorem{problem}{Problem}

\begin{document}

\maketitle
\thispagestyle{empty}
\pagestyle{empty}

\begin{abstract}     
  In this paper, we propose a distance-certified safety control
  framework for unknown and dynamic environments. The proposed
  strategy relies on a spatio-temporal composition of Gaussian Process
  (GP) barriers from environment point clouds.
  A point-cloud-dependent covariance kernel is constructed to smoothly
  combine individual GP representations, yielding a continuously
  differentiable posterior mean and a diagonal observation Gram matrix
  that avoids dense covariance inversions. We then derive a
  closed-form calibration of the GP length scale accounting for
  spatial approximation errors and finite point-cloud coverage,
  guaranteeing a prescribed minimum distance from the true obstacle
  boundary.
  To handle moving obstacles and updated observations, the scan-based
  barriers are certified over finite validity horizons and smoothly
  combined into a time-varying Control Barrier Function (CBF). Under
  problem feasibility conditions, the associated CBF-based safety
  control guarantees forward invariance of the certified safe set, thereby
  ensuring the prescribed distance from the obstacles.
  Unicycle simulations with static and moving obstacles illustrate the
  effectiveness of the proposed approach.
\end{abstract}

\section{Introduction}\label{sec:intro}

The increasing complexity of autonomous systems operating in real-world unstructured environments has led to extensive research in the field of safety-critical control, whose goal is to enforce safety specifications while preserving the performance of a nominal control policy (see, as a reference, the recent survey~\cite{hsu2023safety}).
Among other methods, Control Barrier Functions (CBFs) provide a systematic framework for this purpose by establishing forward invariance of prescribed safe sets \cite{ames2016control,ames2019control}.
These methods have found widespread applications in robotics, including collision avoidance in multi-robot systems~\cite{wang2017safety}, robotic manipulation and human-robot interaction~\cite{ferraguti2022safety,singletary2021safety,landi2019safety}, as well as autonomous vehicles \cite{alan2023control}.
Nevertheless, their implementation requires a suitable mathematical representation of the safe set. In obstacle avoidance problems, such a representation depends on the geometry of the surrounding environment, which is often unknown a priori.

To address this limitation, several approaches have investigated the construction of CBFs from available measurements and/or learned representations.
These include learning barriers from expert demonstrations \cite{robey2020expert}, accounting for perception errors \cite{dean2021perception}, and constructing implicit obstacle representations from range measurements \cite{long2021memory}. 
More directly, point-cloud-based methods have been proposed to construct local barrier functions \cite{desa2024pointcloud}, enable safe navigation in unknown environments \cite{dai2024sailing}, and efficiently compute distance functions for nonconvex obstacles \cite{lutkus2025nonconvex}.

Among other approaches, Gaussian Processes (GPs) provide a flexible framework for representing unknown functions from data, while also quantifying the associated uncertainty \cite{williams2006gaussian}.
Their integration with CBFs has been successfully explored in several settings, including multi-robot collision avoidance \cite{hu2022decentralized}, safe learning in unknown environments \cite{li2023robust}, and online LiDAR-based navigation \cite{keyumarsi2023lidar}.
Further developments have considered GP-based estimation of uncertain dynamics for high-order safety filtering \cite{zhang2024realtime}, and dynamic GP-based barrier functions for moving obstacles \cite{yin2025dynamic}.
Alternative implicit safety representations, such as the Poisson safety function \cite{bahati2025poisson}, have also been proposed. 
Despite these advancements, establishing a direct relationship between the level sets of a GP-based obstacle representation and a user-defined minimum distance to the obstacles remains challenging, especially in the presence of finite boundary sampling.

An additional difficulty arises when the environment evolves over time. Moving obstacles and periodically updated measurements require safety certificates to account for both the evolution of the underlying constraints and the replacement of previously acquired information. 
Time-varying CBFs extend the standard forward-invariance formulation to evolving safe sets \cite{wu2016safety,lindemann2019control,XU2018195}, with applications ranging from environment-adaptive assistance \cite{igarashi2019timevarying} to mobile-robot navigation \cite{huang2023unicycle} and dynamic obstacle avoidance \cite{dai2025differentiable}.
In parallel, smooth compositions of multiple barrier functions have been investigated to accommodate multiple safety requirements \cite{molnar2023composing}. 
Relevant to perception-driven control, a time-varying soft-maximum CBF construction has been proposed in \cite{safari2025time}.
However, the smooth temporal composition of barrier functions does not, by itself, establish a prescribed minimum distance from the true obstacle boundary. 
These considerations highlight two closely related challenges in
perception-based safety-critical control: (i) learning, from finite
boundary sampling, a smooth obstacle representation with
distance-certified level sets, and (ii) providing a distance-aware
safety policy in the presence of moving obstacles.
Addressing both aspects requires connecting the geometry of obstacle
point-cloud observations with time-varying safety control
certificates.

In this paper, we propose a perception-aware safety controller based
on Gaussian Processes with distance-certified guarantees. The
resulting barrier function is guaranteed to maintain a prescribed
minimum distance from the true obstacle boundary, despite finite
point-cloud sampling and obstacle motion.
Specifically, the main contribution of the paper is threefold.
First, we introduce a novel positive-semidefinite covariance kernel that combines individual observations through smooth spatial weights. 
The resulting GP admits an analytical, continuously differentiable posterior mean and a diagonal observation Gram matrix, which avoids dense conditioning inversions and enables efficient online updates.
Second, we establish a deterministic relationship between the GP posterior superlevel set and the complete obstacle boundary.
By deriving a bound on the spatial weighting approximation and accounting for finite point-cloud coverage, we obtain a closed-form calibration of the GP length scale that guarantees a user-defined minimum distance to the true obstacle boundary. We further extend this result to moving obstacles by accounting for bounded obstacle motion and finite scan-validity horizons.
Third, we develop a time-varying safety controller that combines the certified scan-based barriers through smooth temporal weights, and we establish forward invariance of the safe set and a minimum distance to the true obstacle boundary.

The paper is organized as follows. Section \ref{sec:preliminaries} reviews some preliminaries on Gaussian Processes and Control Barrier Functions. 
Section \ref{sec:problem} introduces the considered framework and formulates the problem.
Section \ref{sec:main_result} presents the main theoretical result of the paper and the corresponding algorithmic implementation, while Section \ref{sec:theoretical_analysis} provides the supporting theoretical analysis, including the proof of the main result.
Section \ref{sec:results} evaluates the proposed approach with numerical simulations.
Formal proofs of the supporting theoretical results are provided in the Appendix.

\paragraph*{Notation}
    $\R_{\geq 0}$ and $\mathbb{N}_{\geq 0}$ denote the nonnegative real numbers and nonnegative integers, respectively.
    We denote by $\nabla f$ the gradient of a differentiable function $f$, and by $\partial_t f$ its partial derivative with respect to time $t$.
    A continuous, strictly increasing function $\alpha : \R \to \R$ with $\alpha(0) = 0$ is of extended class $\mathcal{K}$.
    The indicator $\mathbf{1}_{i=j}$ equals $1$ if $i = j$ and $0$ otherwise.
    The floor function $\lfloor r \rfloor$ denotes the greatest integer less than or equal to $r$.
    The Euclidean norm of a vector $x \in \realspace^n$ is denoted by $\|x\|$.
    For a nonempty set $A$, the distance from a point $x$ to the set $A$ is denoted by $\pointsetdist(x,A):=\inf_{a\in A}\|x-a\|$.
    For nonempty subsets $A,B$ of the same Euclidean space, their Hausdorff distance is defined as
    $d_H(A,B) := \max\{\sup_{a \in A} \pointsetdist(a,B), \sup_{b \in B} \pointsetdist(b,A)\}$.
    For any $\tau>0$, define the continuously differentiable, compactly supported window $\chi_\tau(s):= \sin^2\left(\frac{\pi s}{\tau}\right) \mathbf{1}_{[0,\tau]}(s)$.

\section{Preliminaries}\label{sec:preliminaries}

\subsection{Preliminaries on Gaussian Processes}
A Gaussian Process (GP) is a stochastic process defining a distribution over functions~\cite{williams2006gaussian}. We denote a GP by
\begin{align}
    f(x) \sim \mathcal{GP}(\mu(x), k(x, x')),
\end{align}
where $\mu : \R^n \to \R$ is the mean function and $k : \R^n \times \R^n \to \R$ is the covariance function.
The covariance function must be symmetric and positive semidefinite, ensuring that the covariance matrix of any finite collection of function values is positive semidefinite.
A critical property of GPs is that they are closed under conditioning, allowing us to compute the posterior distribution of the function given observed data. 
To this end, consider a dataset $\mathcal{D} = \{(x_i, y_i)\}_{i = 1}^N$ of $N$ input-output pairs, where $y_i = f(x_i) + \epsilon_i$ and $\epsilon_i \sim \mathcal{N}(0, \sigma_n^2)$ is i.i.d. Gaussian noise. 
The posterior distribution of $f$ at a new input $x_*$ is given by
\begin{align}
    \mu_*(\xstar) \!\! &= \! \mu(\xstar) \! + \! k(\xstar, X)(K \! + \sigma_n^2 I)^{-\!1}(y \! -
        \mu(X)), \label{eq:gp_posterior_mean}\\
    {\sigma^2_*(\xstar)} \!\! &= \! k(\xstar, \xstar) - k(\xstar, X)(K + \sigma_n^2 I)^{-\!1}k(X,\xstar), \label{eq:gp_posterior_variance}
\end{align}
where $\mu(X) = [\mu(x_1),\ldots,\mu(x_N)]^\top$ is the vector of prior mean evaluations, $X = [x_1, \dots, x_N]^\top$ is the matrix of training inputs, $y = [y_1, \dots, y_N]^\top$ is the vector of training outputs, and $K \in \R^{N \times N}$ is the Gram covariance matrix with entries $K_{ij} = k(x_i, x_j)$. 
A common choice for the covariance function is the squared-exponential kernel, defined as
\begin{align}
    k(x, x') = \sigma_f^2 \exp\left(-\frac{\|x -
        x'\|^2}{2\ell^2}\right). \label{eq:squared_exponential_kernel}
\end{align}

\subsection{Preliminaries on Control Barrier Functions}\label{subsec:cbf_preliminaries}
Consider the nonlinear control-affine system
\begin{equation}
    \dot \controlstate = f(\controlstate) + g(\controlstate)\controlinput,
    \qquad \controlstate(0) = \controlstate_0,
    \label{eq:plant}
\end{equation}
where $\controlstate(t) \in \cX \subseteq \realspace^{n}$ and $\controlinput(t) \in \cU \subseteq \realspace^{m}$ are the state and control input, respectively.  
The functions $f: \realspace^{n} \to \realspace^{n}$ and $g: \realspace^{n} \to \realspace^{n \times m}$ are locally Lipschitz continuous and $\cU \subset \realspace^{m}$ is nonempty, compact, and convex.
Let $\cbf: \R_{\geq 0} \times \cX \to \R$ be a continuously differentiable function and define the time-varying safe set $\cC(t) := \{x \in \cX : \cbf(t,x) \geq 0\}$.
The family of sets $\cC(t)$ is forward invariant if any closed-loop trajectory initialized at $x(t_0) \in \cC(t_0)$ satisfies $x(t) \in \cC(t)$ for all $t \geq t_0$.
The function $\cbf$ is a time-varying CBF \cite{lindemann2019control,XU2018195} if there exists a function $\alpha \in \cK$ such that
\begin{align}
    \sup_{u \in \cU} \dot \cbf(t, x, u) \geq -\alpha(\cbf(t, x)), \label{eq:tvcbf_condition}
\end{align}
for all $t \geq 0$ and $x \in \cC(t)$, where $\dot \cbf(t, x, u)$ is the total derivative of $\cbf$ along the trajectories of system~\eqref{eq:plant}:
\begin{align}
    \dot \cbf(t, x, u) = \partial_t \cbf(t, x) + \nabla_x \cbf(t, x)^\top (f(x) + g(x)u). \label{eq:tvcbf_total_derivative}
\end{align}
Condition~\eqref{eq:tvcbf_condition} ensures the existence of an admissible input $u \in \cU$ satisfying the CBF inequality, whose enforcement guarantees forward invariance of the safe set $\cC(t)$.
The time-invariant case \cite{ames2016control} is recovered by setting $\cbf(t,x) = \cbf(x)$, so that $\partial_t \cbf(t, x) = 0$ and $\cC(t) = \cC$ for all $t \geq 0$.

\section{Problem Formulation: \\Distance-Certified Safe Control}
\label{sec:problem}
Consider the control-affine system~\eqref{eq:plant}, and let
$\controlmap: \cX \to \workspace$ be a $\cC^2$ map associating the state $\controlstate$ with a point $\workspacepoint = \controlmap(\controlstate)$ in the workspace $\workspace \subseteq \realspace^{n_p}$.
Let $\obstacle(t) \subset \workspace$ denote the closed
set occupied by possibly moving obstacles, with nonempty compact boundary $\trueboundary(t)$.
We first characterize the evolution of the obstacles. 
\begin{assumption}
    \label{ass:obstacle_motion}
    There exist known constants $\velobsmax, \velboundmax \geq 0$,
    such that, for every $s \in [0, t]$,
    \begin{subequations}
        \begin{align}
            \hausdorffdist(\obstacle(t), \obstacle(s)) &\leq \velobsmax(t - s),\label{eq:obstacle_motion_bound}\\
            d_{\mathrm{H}}(\trueboundary(t), \trueboundary(s)) &\leq \velboundmax(t - s),\label{eq:obstacle_boundary_motion}
        \end{align}\label{eq:obstacle_motion}
    \end{subequations}
    where $d_{\mathrm{H}}$ is the Hausdorff distance (cf. Notation).\oprocend
\end{assumption}
Given a prescribed margin $\requiredclearance > 0$, the physical safe set is
\begin{align}
    \safesetphys(t) := \{x \in \cX \mid \pointsetdist(\controlmap(x), \trueboundary(t)) \geq \requiredclearance,  \controlmap(x) \notin \text{int}(\obstacle(t))\}.
    \label{eq:clearance_constraint}
\end{align}
The true obstacle geometry $\obstacle(t)$ is unknown to the controller. Instead, at discrete time instants $t_k = kT$, with $k \in \mathbb{N}$ and sampling period $T > 0$, a perception system (e.g., a camera) provides a finite collection of $\nobs_k$ distinct boundary observations:
\begin{align}
    \obspoints^k := \{\obspoint{1}^k, \ldots, \obspoint{\nobs_k}^k\} \subset \trueboundary(t_k).
    \label{eq:samples}
\end{align}
To account for the discrepancy between the observed point cloud and the complete obstacle boundary, we introduce the following assumption.
\begin{assumption}[complete-boundary coverage]
\label{ass:coverage}
    For every $k \in \mathbb{N}$, the point cloud $\obspoints^k$ (cf.~\eqref{eq:samples}) satisfies
    \begin{align}
        \hausdorffdist(\obspoints^k, \trueboundary(t_k)) \le \covererror^k,
    \end{align}
    for some $\covererror^k > 0$, where $\hausdorffdist$ is the Hausdorff distance. \oprocend
\end{assumption}
\begin{problem}[Distance-Certified Safe Control]
    \label{prob:distance_certified}
    Given system~\eqref{eq:plant}, a nominal control policy $\nominalinput$, a prescribed minimum distance $\requiredclearance > 0$, and a sequence of point-cloud observations $\{\obspoints^k\}_{k \in \mathbb{N}_{\geq 0}}$, design a continuously differentiable time-varying safety function $\safetyfun$ and a safety controller $\safeinput$ such that the certified set 
    \begin{align}
        \cC(t) := \{x \in \cX \mid \safetyfun(t,x) \geq 0\}\label{eq:x}
    \end{align}
    satisfies, for all $t \geq 0$, the following properties:
    \begin{align}
        \safetyfun(t,x) \geq 0
        &\implies
        \pointsetdist(\controlmap(x), \trueboundary(t))
        \geq \requiredclearance,
        \label{eq:problem_clearance}
        \\
        x(0) \in \cC(0)
        &\implies
        x(t) \in \cC(t).
        \label{eq:problem_invariance}
    \end{align}
\end{problem}

\section{Distance-Certified GP-CBF Safe Controller: \\Strategy
  Description and Safety Guarantee}\label{sec:main_result}

\begin{figure}[htpb]
    \centering
    \includegraphics[width=0.6\columnwidth]{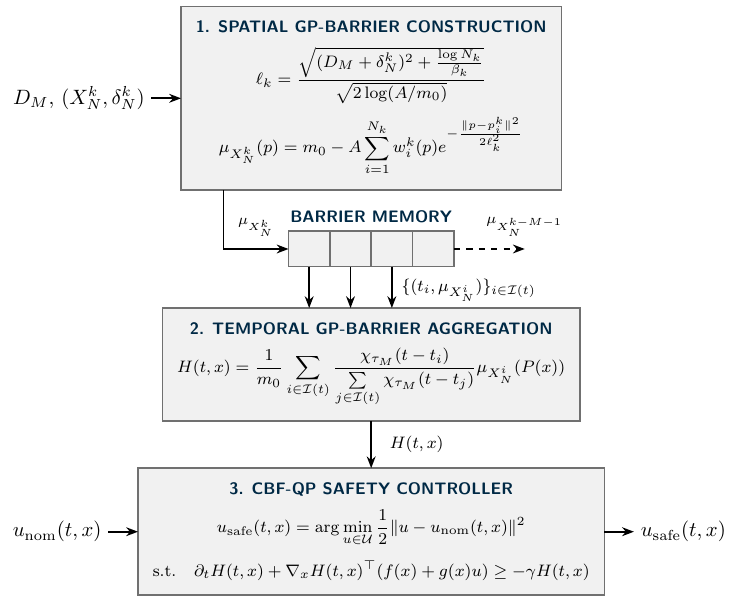}
    \caption{Proposed GP-CBF safety controller structure.}
    \label{fig:gp_cbf_control_scheme}
\end{figure}

To address Problem \ref{prob:distance_certified}, we propose a safety
control strategy that (i) constructs distance-certified GP-based spatial
barriers from successive point-cloud observations, (ii) smoothly aggregates
them into a time-varying safety function, and (iii) enforces the resulting
safety certificate through a CBF-based QP.
The complete procedure is summarized in Algorithm~\ref{alg:gp_cbf}.
    Formally, at each acquisition instant, let $\sqdist{i}^k(p) = \|p - p_i^k\|^2$ denote the squared distance from an observation $\obspoint{i}^k \in \obspoints^k$. 
    We consider a GP with constant prior mean $\prior > 0$, signal variance $\signalvariance > 0$ and a prescribed negative value $\obslabel < 0$ assigned to each observed boundary point.
    Define
    \begin{align}
        \postamp := \frac{(\prior - \obslabel)\signalvariance}{\signalvariance + \noisevariance} \label{eq:alg_post_amp}
    \end{align}
   with $\postamp > \prior$. The spatial weights \eqref{eq:alg_spatial} parametrized by $\beta_k > 0$ and $\selectionpower > \frac{1}{2}$, smoothly combine individual GP representations associated with the observed boundary points $\obspoint{i}^k$ into a single point-cloud-dependent GP representation.
\algrenewcommand\alglinenumber[1]{\scriptsize$\triangleright$}
\begin{algorithm}[H]
\caption{\controlstrategy}
\label{alg:gp_cbf}
\begin{algorithmic}[1]
\State \textbf{Require:} $\nominalinput$, $\postamp$, $\prior$, $\signalvariance$, $\selectionpower$, $\selectionbeta$, $\tau_M$, $D_M$, $d$, $\velboundmax$, $\gamma>0$
\State Acquire initial scan ($X_{N}^0, \covererror^0$) and set $\vartheta_0(0)=1$
\State Initialize $\safetyfun(0,x)$
    \If{a new scan $(X_{N}^k, \covererror^k)$ is available}
        \State Compute spatial weights (with $w_i^k(p_j^k)=\mathbf{1}_{i=j}$)
        \begin{align}
            \weight{i}^k(p) &= \frac{(\sqdist{i}^k(p))^{-\selectionpower} e^{\left(
            -\selectionbeta_k(\sqdist{i}^k(p))\right)}}{\sum_{j = 1}^{\nobs_k}(\sqdist{j}^k(p))^{-\selectionpower} e^{\left(
            -\selectionbeta_k(\sqdist{j}^k(p))\right)}}\label{eq:alg_spatial}
        \end{align}
        \State Set Distance-Certified GP length scale
        \begin{equation}
        \label{eq:alg_calibration}
        \begin{aligned}
        \ell_k &= \frac{\sqrt{(D_M + \covererror^k)^2 + \frac{\log N_k}{\beta_k}}}{\sqrt{2\log (\postamp / \prior)}}
        \end{aligned}
        \end{equation}
        \State Compute GP barrier
        \vspace{-0.7em}
        \begin{align}
        \mu_{\obspoints^k}(p) &= \prior - \postamp \sum_{i = 1}^{\nobs_k} \weight{i}^k(p) e^{\left(-\frac{q_i^k(p)}{2\gplength_k^2}\right)}\label{eq:alg_posterior}
        \end{align}
    \EndIf
        \State Update active scans and temporal weights
        \begin{align}
            \mathcal I(t) &= \{i:0 \leq t-t_i\leq\tau_M\}\label{eq:alg_scans}
        \end{align}
        \begin{equation}
        \label{eq:alg_temporal}
        \vartheta_i(t) = \frac{\chi_{\tau_M}(t - t_i)}{\sum\limits_{j \in \mathcal I(t)} \chi_{\tau_M}(t - t_j)}
        \end{equation}
    \State Compute time-varying barrier function
    \begin{align}
        \safetyfun(t,x) = \frac{1}{\prior}\sum_{i \in \mathcal{I}(t)} \vartheta_i(t) \mu_{\obspoints^i}(\controlmap(x))\label{eq:alg_cbf}
    \end{align}
    \State Compute safe control input
    \begin{align}
    \safeinput(t,x)
    ={}&\arg\min_{u\in\mathcal U}
    \frac12\|u-\nominalinput(t,x)\|^2\label{eq:alg_qp}\\
    \text{s.t.}\quad&
    \partial_t\safetyfun+\nabla_x\safetyfun^\top(f(x)+g(x)u)
    \geq-\gamma \safetyfun\notag
    \end{align}
    \State Apply $\safeinput(t,x)$ to system \eqref{eq:plant}
\end{algorithmic}
\end{algorithm}
    To account for obstacle motion, each scan-based certificate is assigned a validity horizon $\tau_M = (M + 1)T$, with $M \in \mathbb{N}$, and an inflated safety distance $D_M = d + \velboundmax \tau_M$, where $\velboundmax$ is the boundary motion bound in \eqref{eq:obstacle_boundary_motion} and $d$ is the prescribed minimum safety distance we wish to guarantee.
    The GP length scale \eqref{eq:alg_calibration} is calibrated using $D_M$, the point-cloud coverage error $\covererror^k$, and the spatial weighting approximation. 
    To ensure smooth updates, the algorithm retains the active scans \eqref{eq:alg_scans} within the validity horizon $\tau_M$ and combines their associated GP barriers through the temporal weights \eqref{eq:alg_temporal}.
    These are obtained by normalizing the compactly supported window $\chi_{\tau_M}(s)$, defined by specializing the general definition introduced in the Notation paragraph of Section~\ref{sec:intro} to $\tau = \tau_M$.
Unlike the spatial weights, which combine observations within a single scan, the temporal weights regulate the contributions of different scans over time. 
    Since $\chi_{\tau_M}(s)$ and its first derivative vanish at the endpoints of its support, barriers are smoothly introduced and removed. 
    The resulting normalized composite barrier function 
    \begin{equation}
    \label{eq:alg_composite}
    B(t, p) = \frac{1}{\prior}\sum_{i \in \mathcal{I}(t)} \vartheta_i(t) \mu_{\obspoints^i}(p)
    \end{equation}
    is then mapped into the state space through $\controlmap(x)$, yielding the time-varying barrier function $\safetyfun(t,x)$, defined in \eqref{eq:alg_cbf}.
    At every time instant, $\safetyfun(t,x)$ is incorporated into the CBF-QP \eqref{eq:alg_qp}, which computes the safe control input by minimally modifying the nominal policy $\nominalinput(t,x)$ while enforcing the time-varying barrier condition. 
    The overall architecture is also illustrated in Fig.~\ref{fig:gp_cbf_control_scheme}.

    Theorem~\ref{thm:main_result} establishes the corresponding geometric certification and forward invariance guarantees.
    The theoretical analysis supporting these results is presented in Section~\ref{sec:theoretical_analysis}.

\begin{theorem}[Distance-Certified GP-based Safe Controller]\label{thm:main_result}
  Consider system \eqref{eq:plant} and let
  Assumptions~\ref{ass:obstacle_motion} and~\ref{ass:coverage} hold for
  the obstacle model and acquisition system of
  Section~\ref{sec:problem}. Let $\safetyfun(t,x)$ be constructed by
  Algorithm~\ref{alg:gp_cbf} with $d>0$, $\postamp>\prior$,
  $\beta_k>0$, $\selectionpower> \frac{1}{2}$, $M \geq 1$ and
  $\gamma>0$.
    Assume the CBF-QP \eqref{eq:alg_qp} is feasible for all $t\geq 0$ and generates unique, forward-complete closed-loop trajectories for all $x(0)\in\cC(0)$.
    Then, the following hold.
    \begin{enumerate}
        \item \textbf{Geometric Certification:} The barrier $\safetyfun$ is continuously differentiable and satisfies
    \begin{align}
        \safetyfun(t,x) \geq 0 \implies \pointsetdist(P(x),\trueboundary(t)) \geq \requiredclearance,\label{eq:main_result_geometric_certification}
    \end{align}
    for all $t\geq 0$ and $x \in \cX$. 
        \item \textbf{Forward Invariance:} For every $x(0) \in \cC(0)$, the closed-loop system under $\safeinput(t,x)$ satisfies
    \begin{align}
        \safetyfun(t, x(t)) \geq e^{-\gamma t}\safetyfun(0, x(0)) \geq 0,\label{eq:main_result_forward_invariance}
    \end{align}
    ensuring forward invariance of the safe set
    $\cC(t)$. Specifically, if $P(x(0)) \notin \text{int}(\obstacle(0))$, then $x(t) \in \safesetphys(t)$ for all $t\geq 0$.\oprocend
    \end{enumerate}
    The proof of Theorem~\ref{thm:main_result} is provided in Section~\ref{app:proof_theorem_invariance}.
\end{theorem}

\section{Theoretical Analysis}\label{sec:theoretical_analysis}
In this section, we establish the theoretical properties underlying
the proposed approach.
Section \ref{sec:gp_construction_kernel} proves that the barrier
in \eqref{eq:alg_posterior} is the posterior mean of a valid GP
induced by a suitable kernel function, while Section
\ref{sec:closed_form_minimum_distance_certificate} derives the
closed-form minimum-distance certificate.
Section \ref{sec:time_varying_barrier_validity} extends the geometric
guarantee to moving obstacles and characterizes the time-varying
barrier obtained by the temporal combination of stored GP barriers.
Finally, Section \ref{app:proof_theorem_invariance} provides the proof
of Theorem \ref{thm:main_result}, establishing forward invariance of
the GP-based safe set, which guarantees a minimum distance from the true
obstacle boundary.

\subsection{GP Construction and Kernel Validity}\label{sec:gp_construction_kernel}
To analyze the GP construction in Algorithm \ref{alg:gp_cbf}, we
consider a fixed point cloud $\obspoints^k$ at a given acquisition
instant $t_k$.
In the following, we omit the scan index $k$ for notational
simplicity, and we write $\obspoints:= \obspoints^k$,
$\nobs:=\nobs_k$, $\selectionbeta:=\selectionbeta_k$,
$\covererror:=\covererror^k$, and
$\trueboundary:=\trueboundary(t_k)$. The time dependence will be
restored in Section \ref{sec:time_varying_barrier_validity} when
analyzing the temporal combination of GP barriers.
Consider a single boundary observation $\obspoint{i} \in \obspoints$
with label $\obslabel < 0$. For a GP with constant prior mean
$\prior > 0$, squared-exponential kernel as in
\eqref{eq:squared_exponential_kernel}, and length scale
$\gplength > 0$, the single-observation posterior mean obtained from
\eqref{eq:gp_posterior_mean} is
\begin{align}
    \mu_i(p) = \prior - \postamp \exp\left(
    -\frac{\euclidnorm{p - \obspoint{i}}^2}{(2\gplength^2)}
    \right),\label{eq:single_observation_posterior}
\end{align}
where $\postamp$ is defined in \eqref{eq:alg_post_amp}.
Extending this construction to multiple observations is not
straightforward. Indeed, the pointwise minimum of individual posterior
means provides a representation associated with the nearest observed
point, but it is generally nonsmooth where the nearest point
changes. Instead, to obtain the differentiable posterior mean in
\eqref{eq:alg_posterior}, we introduce the smooth spatial weighting
scheme in \eqref{eq:alg_spatial} which satisfies the cardinality
condition $w_i(\obspoint{j}) = \mathbf{1}_{i=j}$.
The corresponding point-cloud-dependent covariance kernel is defined as
\begin{align}
  \gpkernel(p, p^\prime) = \signalvariance e^{\left(
  -\frac{\euclidnorm{p - p^\prime}^2}{(2\gplength^2)}
  \right)} \sum_{i = 1}^{\nobs} \weight{i}(p) \weight{i}(p^\prime). \label{eq:alg_kernel}
\end{align}
Such a kernel allows us to interpret the weighted combination of
individual posteriors in \eqref{eq:alg_posterior} as a
rigorous GP posterior mean, as formalized in the following
result.
\begin{proposition}[Validity of the GP Construction]\label{prop:kernel_validity}
    Consider a fixed point cloud $\obspoints$ and the kernel $\gpkernel$ defined in \eqref{eq:alg_kernel}. Then, $\gpkernel$ is symmetric and positive semidefinite, and its observation Gram matrix satisfies $\gpgram = \signalvariance I_{\nobs}$. 
    Moreover, conditioning the corresponding GP on \eqref{eq:samples} yields the posterior mean defined in \eqref{eq:alg_posterior}. \oprocend
\end{proposition}
The proof of Proposition~\ref{prop:kernel_validity} is provided in Appendix~\ref{app:proof_kernel_validity}.
\begin{remark}
    Thanks to the diagonal structure of the Gram matrix, the GP conditioning does not require the inversion of a dense covariance matrix, and \eqref{eq:alg_posterior} can be expressed as a weighted sum of the individual posteriors in \eqref{eq:single_observation_posterior}. \oprocend
\end{remark}
\begin{remark}
    The geometric certification of the GP superlevel set must account for two sources of approximation: the smooth spatial weighting, which approximates the nearest-observation envelope, and the finite point-cloud sampling, which introduces a discrepancy with respect to the complete obstacle boundary.
    Both effects are accounted for in the calibration of the GP length scale $\gplength$ in \eqref{eq:alg_calibration}, whose minimum-distance guarantee will be established in Lemma \ref{thm:clearance}.
    \oprocend
\end{remark}

\subsection{Closed-Form Minimum-Distance Certificate}\label{sec:closed_form_minimum_distance_certificate}
We now establish that the closed-form calibration of the GP length
scale $\gplength$ in \eqref{eq:alg_calibration} guarantees the
prescribed minimum distance from the true obstacle boundary.
To account for the approximation introduced by the smooth weighting \eqref{eq:alg_spatial}, the calibration incorporates the selection-error bound
\begin{align}
    \selectionerror := \frac{\log\nobs}{\selectionbeta}, \label{eq:selection_error}
\end{align}
which bounds the discrepancy between the weighted and nearest-observation squared distances.
The following lemma establishes the corresponding bound on the weighted Gaussian terms in the posterior mean \eqref{eq:alg_posterior}.
\begin{lemma}
    [Gaussian-Envelope Bound] \label{lemma:selection_error}
    Consider the spatial weights $\weight{i}(p)$ in \eqref{eq:alg_spatial} and let $q_\star(p) := \min_{i}q_i(p) = \pointsetdist(p, \obspoints)^2$. Then, for every $p \in \workspace$ and $\gplength > 0$, it holds
    \begin{align}
         e^{\left(-\frac{q_\star(p) + \selectionerror}{2\gplength^2}\right)} \leq \sum_{i = 1}^{\nobs} \weight{i}(p) e^{\left(-\frac{q_i(p)}{2\gplength^2}\right)} \leq e^{\left(-\frac{q_\star(p)}{2\gplength^2}\right)}. \label{eq:selection_error_bound}
    \end{align}
\end{lemma}
The proof of Lemma \ref{lemma:selection_error} is provided in Appendix \ref{app:proof_lemma_selection_error}.
\begin{figure}[htpb]
    \centering
    \scalebox{0.80}{%
    \begin{minipage}{\columnwidth}
        \centering
        \includegraphics[width=0.98\linewidth]{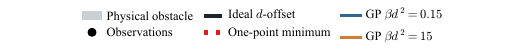}%
        \par\vspace{-1.5mm}%
        \begin{minipage}[t]{0.49\linewidth}
            \centering
            \includegraphics[width=\linewidth]{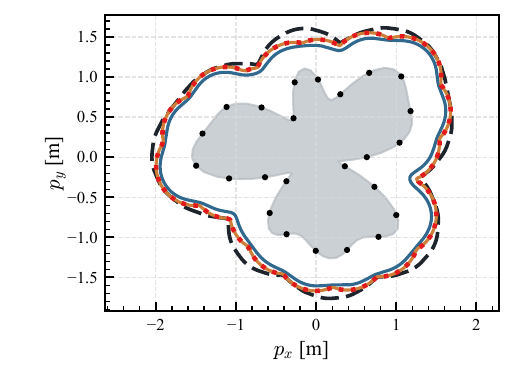}%
            \par\vspace{1mm}%
            \includegraphics[width=\linewidth]{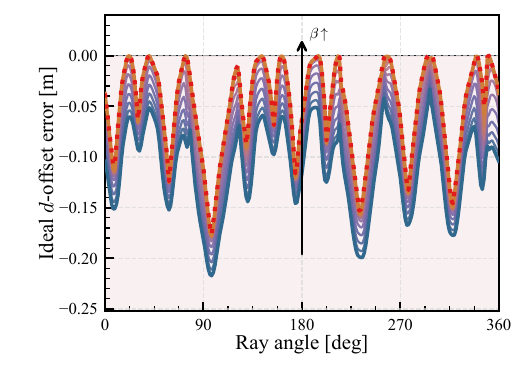}
        \end{minipage}%
        \hfill
        \begin{minipage}[t]{0.49\linewidth}
            \centering
            \includegraphics[width=\linewidth]{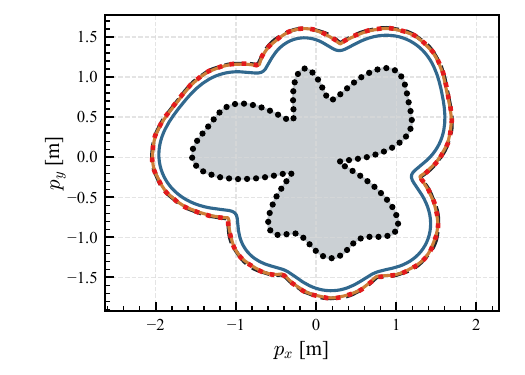}%
            \par\vspace{1mm}%
            \includegraphics[width=\linewidth]{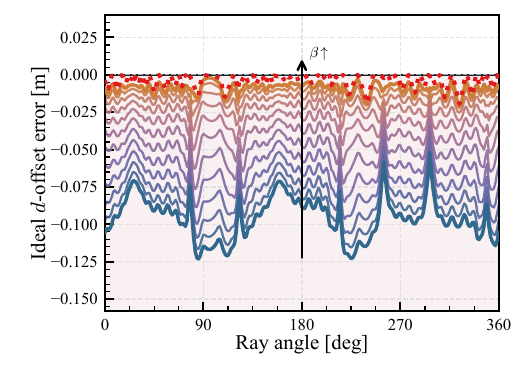}
        \end{minipage}
    \end{minipage}%
    }
    \caption{
        Effect of spatial weighting and boundary sampling on the (uncalibrated) GP zero-level contours for $\nobs = 24$ (left) and $\nobs = 96$ (right) observations. 
        \textbf{Top panels:} (Uncalibrated) GP contours for two values of $\selectionbeta$, compared with the ideal $d$-offset envelope and the one-point minimum envelope.
        \textbf{Bottom panels:} Radial deviations from the ideal $d$-offset for increasing values of $\selectionbeta$. 
    }
    \label{fig:gp_beta_samples}
\end{figure}

    Figure \ref{fig:gp_beta_samples} illustrates the effects of spatial weighting and finite boundary sampling on the GP zero-level contours. 
    For a fixed point cloud, increasing $\selectionbeta$ reduces the selection error bound in \eqref{eq:selection_error}, bringing the GP contour closer to the one-point minimum envelope. 
    The negative deviations shown in the lower panels indicate that the resulting contours lie inside the prescribed minimum distance region.
    These effects motivate the length-scale calibration derived in the following result, which accounts for both spatial weighting and finite boundary coverage.
\begin{lemma}[Minimum-Distance Certificate]\label{thm:clearance}
    Let Assumption \ref{ass:coverage} hold for a point cloud $\obspoints$. Consider the GP posterior $\gpmean$ in \eqref{eq:alg_posterior}. Given any prescribed margin $\requiredclearance > 0$, choosing
    \begin{align}
         \gplength \geq \gplength_{\text{safe}} := \frac{\sqrt{(\requiredclearance + \covererror)^2 + \selectionerror}}{\sqrt{2\log(\postamp/\prior)}}\label{eq:safe_radius}
    \end{align}
    guarantees that, for every $p \in \workspace$,
    \begin{align}
        \gpmean(p) \geq 0 \implies \pointsetdist(p, \trueboundary) \geq \requiredclearance. \label{eq:distance_guarantee}\eqoprocend
    \end{align}
\end{lemma}
The proof of Lemma \ref{thm:clearance} is provided in Appendix \ref{app:proof_lemma_clearance}.

Essentially, \eqref{eq:safe_radius} in Lemma \ref{thm:clearance} provides a closed-form calibration of the GP length scale that guarantees a prescribed minimum distance $\requiredclearance$ from the complete obstacle boundary.
We highlight that choosing $\requiredclearance = D_M$ in \eqref{eq:safe_radius} recovers the calibration of the GP length scale $\gplength$ in \eqref{eq:alg_calibration} adopted in Algorithm \ref{alg:gp_cbf}, extending the guarantee to the case of moving obstacles as will be established in the following subsection.

\subsection{Time-Varying Barrier Construction}\label{sec:time_varying_barrier_validity}
We now extend the minimum-distance certificate derived in Section
\ref{sec:closed_form_minimum_distance_certificate} (cf. Lemma
\ref{thm:clearance}) to the case of moving obstacles by relying on the
temporal combination of stored GP barriers.
We assume that each acquired point cloud is expressed in a common reference frame and kept fixed after acquisition. 
For each scan at time $t_k$, define the normalized GP barrier
\begin{align}
    b_k(p)&=\frac{\mu_{\obspoints^k}(p)}{\prior},\label{eq:alg_posterior_barrier}
\end{align}
and its zero-superlevel set $K_k:=\{p \in \workspace \mid b_k(p) \geq 0\}$.
Recall the validity horizon $\tau_M = (M+1)T$ and the inflated safety distance $D_M = \requiredclearance + \velboundmax \tau_M$ introduced in the algorithm description.
The following result establishes that the calibration \eqref{eq:alg_calibration} preserves the prescribed minimum distance throughout the validity horizon $\tau_M$.
\begin{lemma}
    [Finite-Scan Boundary Validity]\label{lemma:finite_scan_validity}
    Let Assumptions \ref{ass:obstacle_motion} and \ref{ass:coverage} hold.
    Consider the GP barrier $b_k$ constructed in \eqref{eq:alg_posterior_barrier} and its superlevel set $K_k$. Then,
    \begin{align}
        \pointsetdist(p, \trueboundary(t)) \geq D_M - \velboundmax(t - t_k) \geq d
        \label{eq:finite_scan_validity}
    \end{align}
    for all $p \in K_k$ and $t \in [t_k, t_k + \tau_M]$. \oprocend
\end{lemma}
The proof of Lemma \ref{lemma:finite_scan_validity} is provided in Appendix \ref{app:proof_lemma_finite_scan_validity}.
Consequently, every active GP barrier retains its minimum-distance certificate despite obstacle motion. 
This property, together with the smooth temporal composition of stored GP barriers defined in \eqref{eq:alg_composite}, provides the basis for the geometric certification established in Theorem \ref{thm:main_result}.

\subsection{Proof of Theorem \ref{thm:main_result}}\label{app:proof_theorem_invariance}

We first establish the regularity of the composite barrier.
By construction, the spatial weights \eqref{eq:alg_spatial}, with $\selectionpower > \frac{1}{2}$, and the corresponding GP posterior mean \eqref{eq:alg_posterior} are continuously differentiable.
Moreover, since $M \geq 1$ and scans are acquired periodically, the denominator in \eqref{eq:alg_temporal} is strictly positive for all $t > 0$.
As the temporal window $\chi_{\tau_M}$ (cf. Notation) and its derivative vanish at both endpoints of its support $[t_i, t_i+ \tau_M]$, the temporal weights \eqref{eq:alg_temporal} are continuously differentiable across scan updates.
At $t = 0$, the initialization $\vartheta_0(0) = 1$ provides the continuous extension.
Consequently, each stored barrier $b_i$ is continuously differentiable by construction, resulting in the composite barrier $B$ and the corresponding safety function $\safetyfun$ being continuously differentiable.
Since the temporal weights are nonnegative and sum to one, the composite barrier satisfies
\begin{align}
    B(t, p) = \sum_{i \in \mathcal{I}(t)} \vartheta_i(t) b_i(p) \leq \max_{i \in \mathcal{I}(t)} b_i(p).\label{eq:composite_barrier_bound}
\end{align}
Thus, $B(t, p) \geq 0$ implies $b_i(p) \geq 0$ for at least one active scan $i \in \mathcal{I}(t)$. 
Since every active scan lies within its validity horizon, we can apply Lemma \ref{lemma:finite_scan_validity} to guarantee
\begin{align}
    \safetyfun(t,x) \geq 0 \implies \pointsetdist(\controlmap(x), \trueboundary(t)) \geq d.\label{eq:forward_invariance_proof}
\end{align}
The same implication holds at $t = 0$ by the initialization of the composite barrier with the first scan, establishing \eqref{eq:main_result_geometric_certification} and, thus, concluding the proof of the geometric certification of the composite barrier.

Now, consider a closed-loop trajectory initialized in $\cC(0)$. By the assumed feasibility and well-posedness of the CBF-QP, the trajectory exists uniquely for all $t \geq 0$. Applying the chain rule and the CBF condition in \eqref{eq:alg_qp} yields
\begin{align}
    &\frac{d}{dt}\left(
        e^{\gamma t} H
    \right) = e^{\gamma t} \left(
        \partial_t H + \nabla_x H^\top (f(x) + g(x)u) + \gamma H
    \right) \geq 0.
\end{align}
Integrating the above inequality gives
\begin{align}
    H(t, x(t)) \geq e^{-\gamma t} H(0, x(0)) \geq 0,\label{eq:forward_invariance_proof_2}
\end{align}
which establishes \eqref{eq:main_result_forward_invariance} and forward invariance of $\cC(t)$.

Finally, let $p(t) = \controlmap(x(t))$, with $p(0) \notin \text{int}(\obstacle(0))$. Exploiting the obstacle motion bound \eqref{eq:obstacle_motion_bound} (cf. Assumption \ref{ass:obstacle_motion}) and the triangle inequality, we have, for every $t \geq s$
\begin{align}
    \pointsetdist(p(t), \obstacle(t)) &\geq \pointsetdist(p(s), \obstacle(s)) - \|p(t) - p(s)\| - \velobsmax(t - s).\label{eq:triangle_inequality}
\end{align}
By the initial conditions and the condition in \eqref{eq:forward_invariance_proof}, $p(0) \notin \obstacle(0)$. Suppose, by contradiction, that the trajectory $p(t)$ intersects the obstacle, and let $t_\star > 0$ be the first contact time.
For every $s \in [0, t_\star)$, we have
\begin{align}
    \pointsetdist(p(s), \obstacle(s)) = \pointsetdist(p(s), \trueboundary(s)) \geq d > 0.\label{eq:contradiction_1}
\end{align}
Letting $s \to t_\star$, and using continuity of $p(t)$, we obtain $\pointsetdist(p(t_\star), \obstacle(t_\star)) \geq d > 0$.
Applying \eqref{eq:triangle_inequality} again from $s = t_\star$, and using continuity of $p(t)$ we see that the trajectory cannot enter the obstacle for a sufficiently small time interval after $t_\star$, contradicting the assumption that $p(t)$ intersects the obstacle at $t_\star$. 
Thus, $p(t) \notin \obstacle(t)$ for all $t \geq 0$. Together with \eqref{eq:main_result_geometric_certification} this establishes $x(t) \in \safesetphys(t)$ for all $t \geq 0$. \oprocend

\section{Numerical Simulations}
\label{sec:results}
We evaluate the proposed approach on a disk-shaped unicycle robot navigating an unknown environment with static and moving obstacles of different geometries.
\begin{figure}[htpb]
    \centering
    \scalebox{0.80}{%
    \begin{minipage}{\columnwidth}
        \centering
        \includegraphics[width=0.98\linewidth]{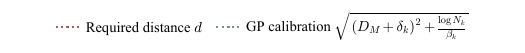}%
        \par\vspace{-1.5mm}%
        \begin{minipage}[t]{0.49\linewidth}
            \centering
            \includegraphics[width=\linewidth]{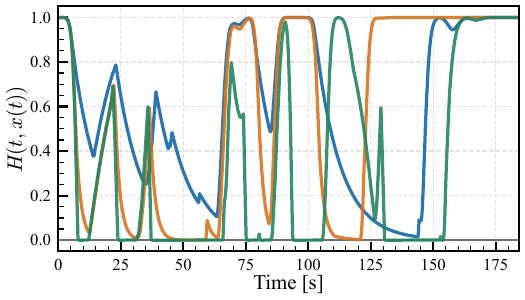}
        \end{minipage}%
        \hfill
        \begin{minipage}[t]{0.49\linewidth}
            \centering
            \includegraphics[width=\linewidth]{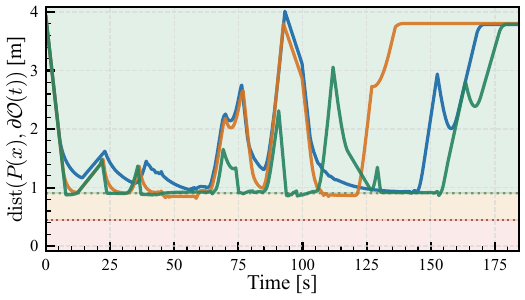}
        \end{minipage}
    \end{minipage}%
    }
    \caption{\textbf{Left:} Time evolution of the GP-CBF $\safetyfun(t,x(t))$ \textbf{Right:}  Protected-point distance for $\gamma=0.1$ (blue), $\gamma=0.5$ (orange), and $\gamma=10$ (green).}
    \label{fig:var_gamma}
\end{figure}
\begin{figure}[H]
    \centering
    \includegraphics[width=0.80\columnwidth, trim=0 0 10.0cm 0.0cm, clip]{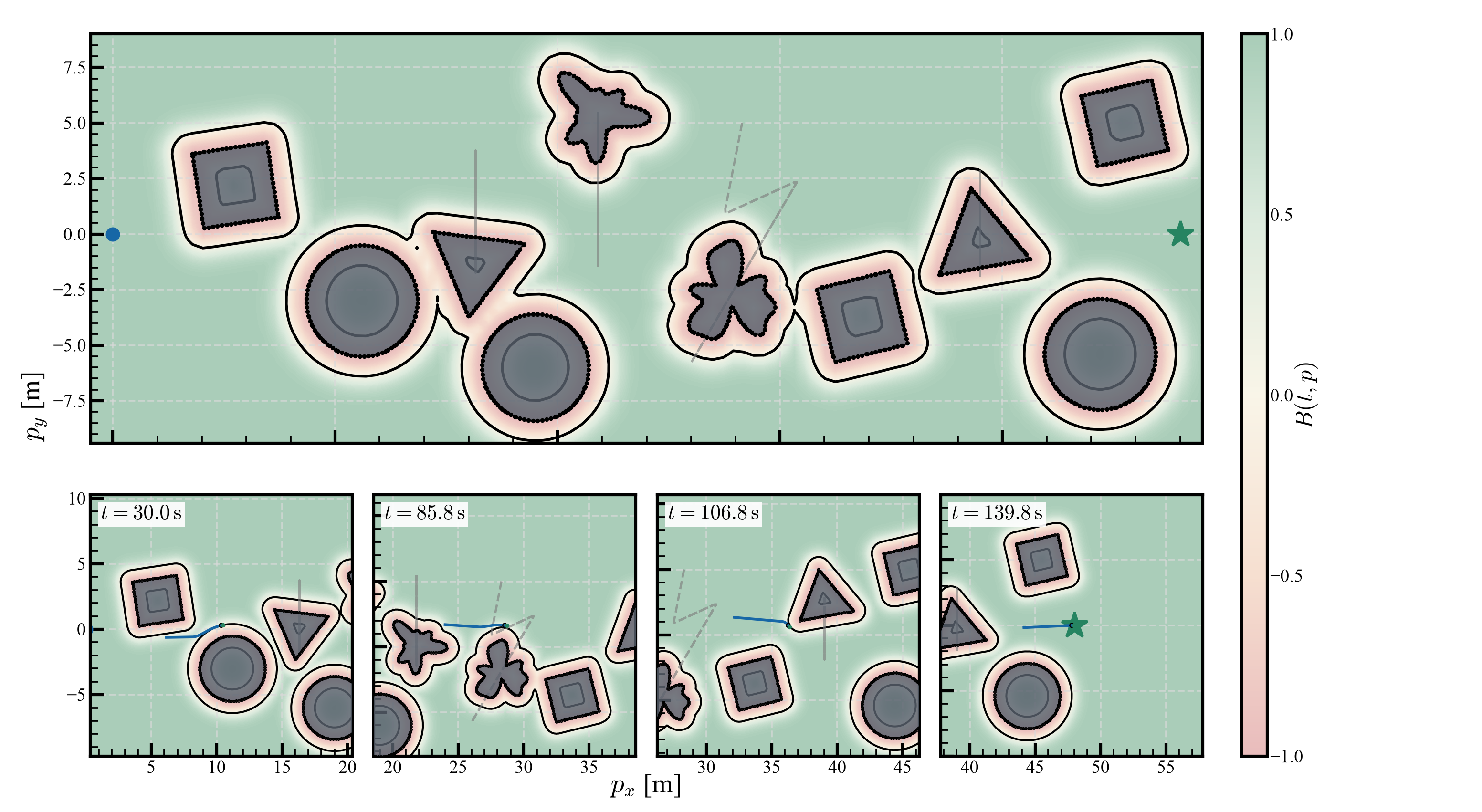}
    \caption{\textbf{Top panel:} Map of the environment. \textbf{Bottom panels:} Snapshots of the dynamic environment and corresponding GP barrier for $\gamma=0.5$. Black dots denote boundary observations, black curves represent the GP zero-level sets, and blue lines show the last $12$ seconds of the robot trajectory.}
    \label{fig:map_frames}
\end{figure}
The robot state is $x = (p_x, p_y, \theta)^\top$, where $(p_x, p_y)^\top$ denotes the robot center position and $\theta$ its orientation. 
The robot dynamics are given by $\dot p_x = v \cos\theta,{}\dot p_y = v \sin\theta,{}\dot\theta = \omega$,
with $|v| \leq 1$ m/s and $|\omega| \leq 6$ rad/s. The robot has radius $r_b = 0.15$ m and protected point $P(x) = (p_x, p_y)^\top + L(\cos\theta, \sin\theta)^\top$, with $L = 0.15$ m. 
A minimum required distance of $d_b = 0.15$ m between the robot body and the obstacle boundary is enforced by prescribing the protected-point distance $d = d_b + r_b + L = 0.45$ m. 
Starting from $x(0) = (0,0,0)$, the robot must steer its lookahead point toward the target point $(48,0)$ m.
The GP parameters are $\prior = 1$, $\signalvariance = 1$, $\selectionpower = 2$, $\selectionbeta = 395$ m$^{-2}$.
Scans are acquired every $T = 0.12$ s, with $M = 6$ and $\tau_M = (M+1)T = 0.84$ s.
To illustrate Assumption \ref{ass:coverage}, we simulate an overhead camera with complete boundary visibility. The workspace is discretized into square cells of side $l_c = 0.20$ m, with each boundary-intersecting cell providing an observation.
This yields a constant coverage bound $\covererror^k = l_c \sqrt{2} \simeq 0.28$ m.
The boundary and obstacle motion bounds are $\velboundmax = \velobsmax = 0.19$ m/s, yielding $\sqrt{(D_M + \covererror)^2 + \selectionerror} = 0.9018$ m.
Since boundary observations are assigned the fixed label $\obslabel = -1$, we set the noise variance to $\noisevariance = 0$.
The nominal controller assigns the protected-point velocity using a workspace proportional controller with $k_g = 0.8$ and maximum velocity $v_g = 0.4$ m/s.
The corresponding physical inputs are obtained through the protected-point Jacobian and saturated at the actuator limits.
The safety controller solves~\eqref{eq:alg_qp} in the scaled input $\widetilde u = (v, L\omega)$. 
We simulate the proposed safety controller for $\gamma \in \{0.1, 0.5, 10\}$.
Trajectories are integrated using the RK45 solver of SciPy, with relative and absolute tolerances $10^{-7}$ and $10^{-9}$, respectively, and a maximum step of $0.02$ s.
Figure~\ref{fig:var_gamma} shows that the proposed strategy maintains nonnegative barrier values and preserves the prescribed minimum distance from the obstacles for all considered values of $\gamma$.
Figure~\ref{fig:map_frames} illustrates the evolution of the GP barrier in the presence of moving obstacles, highlighting the adaptation of the certified safe region to the changing environment.

\section{Conclusions}
We proposed a distance-certified safety control framework for unknown and dynamic environments based on GP barriers constructed from point-cloud observations.
We introduced a novel covariance kernel that enables efficient computation of GP-based safety certificates. 
By accounting for spatial approximation errors and finite point-cloud coverage, we derived a closed-form calibration of the GP length scale that guarantees a prescribed minimum distance from the time-varying obstacle boundaries. 
The resulting safety controller guarantees forward invariance of the certified safe set and preservation of the prescribed distance. 

\appendix

\subsection{Proof of Proposition \ref{prop:kernel_validity}}\label{app:proof_kernel_validity}
Consider $N$ independent zero-mean Gaussian processes $G_i \sim \mathcal{GP}(0, k_{\text{SE}})$, $i = 1, \ldots, N$, with $k_{\text{SE}}$ as in \eqref{eq:squared_exponential_kernel}.
Construct the random field
\begin{align}
    F(p) := \prior + \sum_{i=1}^{N} w_i(p)G_i(p).\label{eq:random_field}
\end{align}
Since the weights are deterministic and the $G_i$ are independent, $F$ is a GP with mean $\prior$ and covariance $\mathrm{Cov}(F(p),F(p^\prime))=k_{\mathrm{SE}}(p,p^\prime)\sum_{i=1}^{N}w_i(p)w_i(p^\prime)=k_{\obspoints}(p,p^\prime)$.
Thus, $k_{\obspoints}$ is symmetric and positive-semidefinite.
Moreover, exploiting the cardinality property of the spatial weights \eqref{eq:alg_spatial}, we obtain
\begin{align}
    k_{\obspoints}(p_j, p_k) &= \signalvariance \sum_{i=1}^{N} w_i(p_j)w_i(p_k) = \signalvariance \mathbf{1}_{j=k},\label{eq:gram_matrix}\\
    k_{\obspoints}(p, p_j) &= \signalvariance w_j(p) e^{-\frac{\sqdist{j}(p)}{2 \gplength^2}}.\label{eq:cross_covariance} 
\end{align}
The equality \eqref{eq:gram_matrix} implies $K_{\obspoints} = \signalvariance I_N$. Substituting both \eqref{eq:gram_matrix} and \eqref{eq:cross_covariance} into the GP conditioning formula (cf. \eqref{eq:gp_posterior_mean}), with $X = [p_1,\ldots,p_N]^\top$ and all outputs equal to $\obslabel < 0$, yields
\begin{align}
    \gpmean(p) = \prior + \frac{\obslabel - \prior}{\signalvariance + \noisevariance} \sum_{i=1}^{N} k_{\obspoints}(p, p_i) = \prior - \postamp \sum_{i=1}^{N} w_i(p) e^{-\frac{\sqdist{i}(p)}{2 \gplength^2}},\label{eq:posterior_proof}
\end{align}
which recovers \eqref{eq:alg_posterior} and completes the proof. \oprocend

\subsection{Proof of Lemma \ref{lemma:selection_error}}\label{app:proof_lemma_selection_error}
At observation points $p_i \in \obspoints$, the claim follows directly from the cardinality property $\weight{i}(p_j) = \mathbf{1}_{i=j}$.
Otherwise, fix $p \in \workspace \setminus \obspoints$ and define the following auxiliary quantities
\begin{align}
    \Delta_i := q_i - q_\star, \quad r_i:= \left(\frac{q_\star}{q_i}\right)^\selectionpower, \quad Z := \sum_{i=1}^{\nobs} r_i e^{-\beta \Delta_i}.\label{eq:auxiliary_quantities}
\end{align}
Since $q_i \geq q_\star > 0$ for all $i$, we have $r_i \in (0,1]$. Moreover, at least one $i_\star$ satisfies $q_{i_\star} = q_\star$, implying $r_{i_\star} = 1$ and $\Delta_{i_\star} = 0$; hence $Z \geq 1$.
The weights $\weight{i}(p)$ in \eqref{eq:alg_spatial} can be thus rewritten as $\weight{i}(p) = \frac{r_i e^{-\beta \Delta_i}}{Z}$.
Taking logarithms and introducing the Shannon entropy $\mathrm{H}(\weight{}) := -\sum_{i=1}^{\nobs} \weight{i} \log \weight{i}$, we obtain
\begin{align}
    \beta \sum_{i=1}^{\nobs} \weight{i} \Delta_i \!=\! \mathrm{H}(\weight{}) \!+\! \sum_{i=1}^{\nobs} \weight{i} \log r_i \!-\! \log Z \overset{(a)}{\leq} \log N,\label{eq:entropy_bound}
\end{align}
where in $(a)$ we use $r_i \leq 1$ and $Z \geq 1$, and the known upper bound on the entropy of a discrete probability distribution over $N$ elements.
Also, since $z \mapsto e^{- z  / (2 \gplength^2)}$ is convex and decreasing,
\begin{align}
    e^{-\frac{q_\star + \selectionerror}{2 \gplength^2}} \overset{(a)}{\leq} e^{-\frac{\sum_i w_iq_i}{2 \gplength^2}} \overset{(b)}{\leq} \sum_{i=1}^N w_i e^{-\frac{q_i}{2 \gplength^2}} \overset{(c)}{\leq} e^{-\frac{q_\star}{2 \gplength^2}},\label{eq:jensen_bound}
\end{align}
where $(a)$ follows from \eqref{eq:entropy_bound} and the monotonicity of the exponential, in $(b)$ we use Jensen's inequality, and $(c)$ follows from $q_i \geq q_\star$ and $\sum_i w_i(p) = 1$. \oprocend

\subsection{Proof of Lemma \ref{thm:clearance}}\label{app:proof_lemma_clearance}
Let $p \in \workspace$ satisfy $\gpmean(p) \geq 0$. Recall that $q_\star(p) = \pointsetdist(p, \obspoints)^2$. Combining the Gaussian-envelope bound \eqref{eq:selection_error_bound} with the definition of the GP posterior mean \eqref{eq:alg_posterior}, we get
\begin{align}
    e^{-\frac{q_\star(p) + \selectionerror}{2 \gplength^2}} \leq \sum_{i=1}^{\nobs} \weight{i}(p) e^{-\frac{q_i(p)}{2 \gplength^2}} \leq \frac{\prior}{\postamp}.\label{eq:gp_mean_bound}
\end{align}
Taking logarithms and using the condition \eqref{eq:safe_radius} yields
\begin{align}
    q_\star(p) \geq 2 \gplength^2 \log(\postamp / \prior) - \selectionerror \geq (d + \covererror)^2.\label{eq:q_star_bound}  
\end{align}
Therefore, by Assumption \ref{ass:coverage}, 
\begin{align}
    \pointsetdist(p, \trueboundary) \geq \sqrt{q_\star} - \covererror \geq d + \covererror - \covererror = d,\label{eq:clearance_bound}
\end{align}
which concludes the proof. \oprocend

\subsection{Proof of Lemma \ref{lemma:finite_scan_validity}}\label{app:proof_lemma_finite_scan_validity}
Fix $k \geq 0$ and let $p \in K_k$. By definition \eqref{eq:alg_posterior_barrier}, $b_k(p) \geq 0$ implies $\mu_{\obspoints^k}(p) \geq 0$. 
Since the calibration \eqref{eq:alg_calibration} corresponds to \eqref{eq:safe_radius} with $d = D_M$, Lemma \ref{thm:clearance} implies
\begin{align}
    \pointsetdist(p, \trueboundary(t_k)) \geq D_M.\label{eq:finite_scan_validity_proof_1}
\end{align}
Moreover, for any $t \in [t_k, t_k + \tau_M]$ 
\begin{align}
    \pointsetdist(p, \trueboundary(t)) &\overset{(a)}{\geq} \pointsetdist(p, \trueboundary(t_k)) - \velboundmax(t - t_k) \overset{(b)}{\geq} D_M - \velboundmax (t - t_k) \overset{(c)}{\geq} d,\label{eq:finite_scan_validity_proof}
\end{align}
where $(a)$ follows from Assumption \ref{ass:obstacle_motion}, $(b)$ follows from \eqref{eq:finite_scan_validity_proof_1}, $(c)$ follows from both the definition of $D_M$ in the algorithm description, and the inequality $t - t_k \leq \tau_M$.
Since $p \in K_k$ was arbitrary, the bound \eqref{eq:finite_scan_validity_proof} holds for all $p \in K_k$. \oprocend

\end{document}